
\input psfig
\magnification 1200
\rightline{RU-94-95}
\rightline{hep-th/9412169}

\vskip 0.4in
\centerline{\bf Exact Partition Function and Boundary State of Critical
Ising Model}
\centerline{\bf with Boundary Magnetic Field}
\vskip 0.6in
\centerline{R.Chatterjee \footnote{$^\dagger$} {email:
robin@physics.rutgers.edu}}

\centerline{Department of Physics and Astronomy}
\centerline{Rutgers University}
\centerline{P.O.Box 849, Piscataway, NJ 08855-0849}
\vskip 0.4in
\centerline{\bf Abstract}

We compute the exact partition function of the 2D Ising Model at
critical temperature but with nonzero magnetic field at the boundary.
The model describes a renormalization
group flow between the free and fixed conformal boundary conditions in
the space of boundary interactions.
For this flow the universal ground state degeneracy $g$ and the full
boundary state is computed exactly.

\vskip 0.4in

\noindent{\bf 1. Introduction}
\vskip 0.1in

A conformal field theory (CFT) [1] is understood to be the ultraviolet
(UV) limit of a
class of general quantum field theories (QFT). It is associated with
fixed manifolds of the renormalization group (RG) flow in general QFT.
And general QFT's are understood as deformations of CFT's and describing RG
trajectories of their corresponding unstable manifolds. More precisely,
they are interpreted as appropriate relevant perturbations of
appropriate CFT's [2]. For this interpretation to be viable, it should be
possible, at least in principle, to compute infrared behavior of a QFT
from its UV definition. In 2D, for certain ``integrable perturbations''
of CFT's, this is made possible by the fact that the scattering (S)
matrix is exactly calculable [2].

The S matrix is understood to encode all information about its
underlying QFT. One
partially successful way of reconstructing the QFT from the S matrix is
the Thermodynamic Bethe Ansatz (TBA) approach [3-6]. In particular, the
method yields a $c$ function, seemingly having all the characteristics
of Zamolodchikov's $c$ function [7].

All the above understanding has been and is still being extended to
QFT with boundary.
CFT on manifolds with boundary is studied in [8-10]. It is shown in [9]
that, corresponding to each conformal family in the (bulk) CFT, there is
a boundary condition which preserves conformal symmetry. Corresponding
to each such conformal boundary condition, there is a boundary state
containing all information about the boundary condition. This boundary
state belongs to the Hilbert space of states in the Hamiltonian picture
based on a spacetime coordinatization in which the boundary is a spacelike
curve. In [9] it is shown how to classify and compute boundary states.

The understanding of CFT with boundary is extended to integrable field
theory on manifolds with boundary in [11]. It is shown that, for a model
which is integrable in the absence of boundary, when defined on
manifolds with boundary, there are certain boundary conditions which
preserve the integrability of the model. The notion of integrable
boundary condition is made precise in [11] and a general theory to
understand and compute exact boundary states and boundary scattering
matrices for
reflections off boundaries is formulated. Boundary scattering
matrices for several models have since been computed [11-16].

Another aspect of boundary QFT and critical phenomena is the universal
ground state degeneracy factor $g$ [17]. In [9] it is shown how to
calculate $g$ for boundary CFT's. In [17], $g$ is conjectured to be
monotonically decreasing from a less stable to a more stable fixed point
along the RG flow of boundary interactions. However as of now, there
exist no fundamental definition of the ``$g$ function'' in terms of
fundamental quantities of the boundary QFT like the correlation
functions, and no proof for the ``$g$ theorem'' conjecture.

In [19] and [20], the Thermodynamic Bethe Ansatz (TBA) method has been used to
compute the boundary contribution to the free energy of boundary
integrable models. However, because of certain difficulties discussed in
these references, it has so far been possible to compute only ratios of
the $g$ function using the TBA approach.

One of the simplest conformally invariant boundary models is the Ising
model with boundary, studied in [8-10]. In [9] the two essentially
different conformally invariant boundary conditions are identified to
be (1) free case: boundary spins are free; and (2) fixed case: all
boundary spins are fixed to $+1$ or $-1$.

In this paper, we study more general boundary conditions breaking
conformal invariance: the Ising spins interacting with an
external  magnetic field at the boundary. The model is solved on the
lattice in [21,22] (see also [23]). In this paper, we demonstrate a
short cut way to
calculate the exact partition function. In addition, we compute the
boundary state. We compute the
partition function on a long cylinder of length $L$ in one Hamiltonian
representation and transform it to the cross channel representation.
This enables us to compute the full boundary state exactly. In
what follows, we describe the model, elaborate on this strategy
and explicitly compute the partition function and boundary state.
\vskip 0.4in

\noindent{\bf 2. Ising Model With Boundary Magnetic Field}
\vskip 0.1in

We study the Ising model with its boundary spins interacting with a
constant external magnetic field $H_B$. This boundary condition
interpolates between the free case when $H_B=0$ and the fixed to $\pm 1$
case when
$H_B\to\pm\infty$. It thus drives the system from the free case in the
UV down to the fixed case in the IR, which are the fixed points of this
boundary
RG flow [17]. The model is studied in [11,18]. At critical
temperature, the continuum limit
of the model is described by the action:

$${\cal A}_{h} = {1\over{2\pi}}\int_{\cal D}d x d y[\psi{\partial_{\bar
z}}\psi + {\bar \psi}{\partial_z}{\bar \psi}] + \int_{\cal
B} dt[-{i\over{4\pi}}\psi{\bar \psi} + {1\over 2}a{\dot a}] + ih\int_{\cal
B}d t\ a(t)(e^{1\over2}\psi + {\bar e}^{1\over2}{\bar \psi})(t)
\eqno(1)$$
Here, $z=x+{i y}$, $\bar{z}=x-{iy}$ are complex coordinates on our
manifold ${\cal D}$; the boundary ${\cal B}$ is given by a parametric
curve ${\cal B}:\ z=Z(t)\ ,\ {\bar z}={\bar Z}(t)$, $t$ being a real
parameter; $e(t)={d\over dt}Z(t)\ ,\ {\bar e}(t)={d\over dt}{\bar
Z}(t)$ are components of the tangent vector $(e,\ {\bar e})$ to the
boundary with $e(t){\bar e}(t)=1$; $\psi, \bar{\psi}$ are free massless
Majorana fermion fields;
$a(t)$ is auxiliary
``boundary'' fermionic degree of freedom (see [11,18]) with the two
point function:

$$\langle a(t)\ a(t')\rangle_{free} = {1\over2}sign(t-t')\quad , \eqno(3)$$
and $\dot a \equiv
{d\over dt}a$; $h$ is the boundary coupling constant (appropriately
rescaled external field $H_B$) with the dimension of
$[Length]^{-{1\over2}}$. The first two terms in (1) describe the
conformally invariant critical Ising model with free boundary
conditions. The third boundary perturbation term is the boundary spin
operator (see [11]):

$$\sigma_B(t) = i a(t)[e^{1\over2}\psi + {\bar e}^{1\over2}{\bar \psi}](t)
\eqno(2)$$
The boundary spin operator was identified in [9,10]. Thus (1) can be
interpreted as the free boundary Ising field theory perturbed by the
boundary spin operator $\sigma_B$.

It is easily seen from (1) that the fermion fields $\psi$, ${\bar \psi}$
satisfy the following boundary condition at each boundary:

$$({d\over{d t}} + i\lambda)\psi(t) = ({d\over{d t}} - i\lambda){\bar
\psi}(t) \eqno(4)$$
where

$$\lambda = 4\pi h^2 \eqno(5)$$
and

$$\psi(t)= e^{1\over2}(t)\psi\big(Z(t)\big), \qquad {\bar \psi}(t)= {\bar
e}^{1\over2}(t){\bar \psi}\big({\bar Z}(t)\big) \eqno(6)$$
In particular, from (6), we see that the free ($h=0$) and fixed ($h\to
{\pm \infty}$) boundary conditions correspond to:

$$\eqalign{\psi(t) = {\bar \psi}(t) \hbox{, \qquad free case,}\cr
\psi(t) = -{\bar \psi}(t) \hbox{, \qquad fixed case}} \eqno(7)$$
At the left ($x=0$) and right ($x=L$) boundaries of the cylinder (see
Fig.1) the
boundary condition (4) takes the forms:

$$({d\over dy}+ i\lambda){\bar\omega}\psi(-i y)= ({d\over dy}-
i\lambda)\omega{\bar\psi}(i y)\hbox{\qquad at left boundary, \ and} \eqno(8a)$$

$$({d\over dy}+ i\lambda)\omega\psi(L+ i y)= ({d\over dy}-
i\lambda){\bar\omega}{\bar\psi}(L- i y) \hbox{\qquad at right boundary,}
\eqno(8b)$$
where $\omega=e^{i\pi\over4}$ and ${\bar\omega}=e^{-i\pi\over4}$.
\vskip 0.4in

\noindent{\bf 3. Method Outline}
\vskip 0.1in

Quantum Field Theory on the space-time of a cylinder of length $L$
(later we will take $L$ to be large) and
circle length $R$ (see Fig.1), coordinatized by $x$ and $y$ respectively,
can be viewed in two alternate Hamiltonian pictures: In the first, one
considers  the (Euclidean)
time coordinate to run along the length of the cylinder (i.e. $x$) and the
space coordinate along the circle (i.e. $y$). A Hilbert space of states
${\cal H}_R$ is associated with any time slice $x=const$  across the cylinder
with fields $\psi$, ${\bar \psi}$ satisfying periodic (for Ramond sector)
or antiperiodic (for
Neveu-Schwarz sector) boundary condition along
the circle. Naturally, ${\cal H}_R$ is the same as in the bulk theory
since in this picture there is no spatial boundary. The partition
function is given by:

$${\cal Z}=\langle B_{right}|e^{-L H_R}|B_{left}\rangle \eqno(9)$$
where $|B_{left}\rangle$ and $|B_{right}\rangle$ are the boundary
states corresponding to the boundary conditions at the $x=0$
and $x=L$ ends of the cylinder respectively, and $H_R$ is the
Hamiltonian in this picture.

In the other picture,
the space and time coordinates are interchanged (i.e. $x$ becomes the
space coordinate and $y$ the time coordinate) and we now have a
system of particles on a line of length $L$ at temperature $T={1\over R}$.
The space of states ${\cal H}_L(B_{left},B_{right})$ is  associated with a
$y=const$ curve and
these states must now satisfy boundary conditions $B_{left}$,
$B_{right}$ at $x=0$ and
$x=L$ cylinder ends respectively. The partition function is given by

$${\cal Z_{\pm}}=tr((\pm 1)^F e^{-R H_L})\eqno(10)$$
where the $+$ and $-$ signs correspond to antiperiodic and periodic
boundary conditions along $y$, $F$ is the fermion number operator and $H_L$
is the Hamiltonian in this picture. In this picture the free fields
$\psi$, ${\bar\psi}$ admit the following decomposition in terms of plane
waves:

$$\psi(z)= \sum_{k_l}{b_{k_l}e^{i k_l z}} \qquad,\qquad {\bar\psi}=
\sum_{k_l}{{\bar b}_{k_l}e^{-i k_l{\bar z}}} \eqno(11)$$
where the sum is over the free particle momenta $k_l$ which are
constrained to satisfy:

$$1 + X(k_l) = 0 \hbox{\quad ,\qquad where\qquad} X(k)= e^{2 i k_l
L}\Big({{k- i\lambda}\over{k+ i\lambda}}\Big)^2 \eqno(12)$$
where (12) is obtained by imposing (8a) and (8b) on (11). In (11),
$b^{\dagger}_{k_l}=b_{-k_l}$ (notice that solutions of (12) occur
in pairs $\pm k_l$) and $b_{k_l}$, $b^{\dagger}_{k_l}$ with $k_l > 0$
($k_l$=0 does
not satisfy (12)) are the usual free fermion
creation and annihilation operators
satisfying standard anticommutation
relations $\{b_{k_n} , b^{\dagger}_{k_m}\}= \delta_{k_n , k_m}$\ and
$\{b_{k_n} , b_{k_m}\}=0= \{b^{\dagger}_{k_n} , b^{\dagger}_{k_m}\}$.
And, from (8a),
${\bar b}_{k_l}= -i({k_l +i\lambda\over k_l - i\lambda})b_{k_l}$,
\ implying that ${\bar b}_{k_l}s$ do not represent independent degrees of
freedom.

In the next sections we start with the partition function as sum
over the states in (10) (described by modes in (11),(12)). We want to
transform it so that it represents the partition function (9) in the
cross channel picture. {}From this we can compute the boundary state.
\vskip 0.4in

\noindent{\bf 4. Partition Function}
\vskip 0.1in

The Hamiltonian of our model in the finite temperature field theory
picture ($x$ is space coordinate and $y$ is time coordinate) is:

$$H_L= E_L + \sum_{k_l>0}k_l\ b_{k_l}^{\dagger}\ b_{k_l} \eqno(13)$$
where ${k_l}^{'}s$ are the
allowed momenta
satisfying (12), and $E_L$ is the
ground state energy. The partition function (9) is:

$${\cal Z}_{\pm} = e^{-R E_L}\prod_{k_l>0}\bigl(1\pm e^{-R k_l}\bigr)
\eqno(14) $$
Here, as before, the $\pm$ signs refer to antiperiodic and periodic
boundary conditions along the cylinder circle satisfied by
the fields $\psi, \bar{\psi}$. At thermodynamic equilibrium the free energy
$F_{\pm}$ is given by:

$$-R F_{\pm} = -R E_L + \sum_{k_l>0}\log\bigl(1 \pm e^{-R k_l}\bigr)
\eqno(15) $$
This can be written as:

$$-R F_{\pm} = -R E_L + {1\over2\pi i}\int_{\cal C}dk {X'\over{1 +
X}}\log(1 \pm e^{-R k}) \eqno(16) $$
where the integration contour ${\cal C}={\cal C_+}+{\cal C_-}$ in the
complex $k$ plane and the positions of poles of the integrand are shown
in Fig.2. In the above, $X'={d X \over{d k}}$. We write the
integral along contour ${\cal C_-}$ as

$$\int_{\cal C_-}{dk{X'\over{1+X}}\log(1\pm Y_R)}=\int_{\cal
C_-}{dk{X'\over X}\log(1 \pm Y_R)} - \int_{\cal
C_-}{dk{X'\over{X^2(1+{1\over X})}}\log(1\ \pm Y_R)} \eqno(17) $$
where $Y_R = e^{-k R}$. Now we change the integration variable $k\to -k$ in
the second integral on the RHS of (17), and noting that

$$X(-k)={1\over X(k)} \qquad,\qquad Y_R(-k)={1\over Y_R(k)}\quad , \eqno(18)$$
we get:

$$\eqalign{{1\over{2\pi i}}\int_{\cal C_-}{dk{X'\over{1+X}}\log(1\pm Y_R)}=
{1\over{2\pi i}}\int_{\cal C_-}{dk{X'\over X}\log(1\pm Y_R)}\ +\cr
+\ {R\over{2\pi i}}\int_{-{\cal C_-}}{dk {k X' \over 1+X}\log(1\pm Y_R)} +
{\nu_{\pm}\over 2}\log2} \eqno(19) $$
Here, the contour ${\cal -C_-}$ is shown in Fig.3; $\nu_+=0$ and
$\nu_-=1$. The last
term in (19) arises as a result of writing $\log(Y_R - 1)$ as analytic
continuation of $\log(1 - Y_R)$ and we have used

$$\int\limits_{-{\cal C_-}}{dk{X'\over1+X}}= \log(1+X)\Bigm\vert_0^{i\infty}=
-\log2 \eqno(20)$$
{}From (16), (19) we get:

$$\eqalign{-R F_{\pm}= -R E_L + {1\over{2\pi i}}\int_{\cal
C'}{dk{X'\over1+X}\log(1\pm Y_R)}\ + \cr
+\ {1\over\pi}\int\limits_0^\infty{dk(L
+ {2\lambda\over{k^2 + \lambda^2}})\log(1\pm Y_R)}
+{R\over 2\pi
i}\int\limits_0^{i\infty}{dk k{X'\over1+X}} + {\nu_{\pm}\over2}\log2}
\eqno(21) $$
Here ${\cal C'}={\cal C}_+ + -{\cal C_-}$ is shown in Fig.3. Now,

$$H_L={1\over2}\sum_{k_l>0}{k_l(b_l^{\dagger}b_l - b_lb_l^{\dagger})}=
\sum\limits_{k_l>0}(b_l^{\dagger}b_l - {k_l\over2}) \eqno(22)$$
which when compared with (13) implies
$E_L=-{1\over2}\sum\limits_{k_l>0}k_l$. This, as before, can be written
as:

$$E_L= -{1\over{4\pi i}}\int_{\cal C}{dk k{X'\over1+X}} \eqno(23) $$
which yields the following:

$$E_L= {1\over{2\pi i}}\int\limits_0^{i\infty}{dk k{X'\over1+X}} -
{1\over2\pi}\int\limits_0^{\infty}{dk k(L +
{2\lambda\over{k^2+\lambda^2}})} \eqno(24) $$
In (23), the integration contour is the same as in (16). In (24), the
two terms in the integral in the RHS are the usual bulk and
boundary intensive free energies respectively.
We denote the
boundary term as $\varepsilon_\lambda$:

$$2\varepsilon_\lambda=
-{1\over\pi}\int\limits_0^{\infty}{dk{2\lambda k\over{k^2+\lambda^2}}}
\eqno(25) $$
$2\varepsilon_\lambda$ goes as $\lambda\log(\lambda^2) +$
nonuniversal
terms which we set to $0$. Likewise, we set the nonuniversal term in
(24)
${1\over2\pi}\int_0^{\infty}{dk k}$ to zero. In (21), we evaluate the integral
in the second term
by parts and residues and get:

$$\eqalign{{1\over{2\pi i}}\int_{\cal C'}{dk{X'\over1+X}\log(1\pm Y_R)}=
-{1\over{2\pi i}}\int_{\cal C'_W}{dk{\pm Y_R'\over{1\pm
Y_R}}\log(1+X(k))} \cr
=\left\{\eqalign{\log\Sigma_+&=\sum\limits_{l\ge 0}{\log(1+X(i
\omega_l))},\qquad
\omega_l={2\pi l\over
R},\qquad l\in {{\bf Z}+{1\over2}},\qquad\hbox{for} (+)\cr
\log\Sigma_-&=\sum\limits_{n>0}{\log(1+X(i \omega_n))},\qquad
\omega_n={2\pi n\over R},\qquad n\in {\bf Z},
\qquad\hbox{for} (-)\cr}\right\}} \eqno(26)$$
where the Wick rotated contour $\cal C'_W$ is shown in Fig.4. Putting
(21), (24), (26) together, we get:

$$\eqalign{-R F_{\pm}= -2 R\varepsilon_\lambda + \log\Sigma_{\pm} +
{L\over\pi}\int\limits_0^{\infty}{dk\log(1\pm Y_R)}\ +\cr
+\ {1\over\pi}\int\limits_0^{\infty}{dk{2\lambda\over{k^2+\lambda^2}}\log(1\pm
Y_R)} + {\nu_{\pm}\over2}\log2\cr} \eqno(27) $$
In (27), we recognize the third term to be the Casimir
energy on the circle. The $(+)$ sign corresponds to the Neveu-Schwarz
sector and $(-)$ sign to the  Ramond sector.
We denote these by $E_R^{(\pm)}$:

$$E_R^{(\pm)}={-1\over\pi}\int\limits_0^{\infty}{dk\log(1\pm e^{-k R})}
\eqno(28)$$
The fourth and fifth terms in (27) are to be identified as related with
the ground state degeneracy factor $g(R)$.

$$\log g_{\pm}(R)=
{\lambda\over\pi}\int\limits_0^{\infty}{dk{1\over{k^2+\lambda^2}}\log(1\pm
e^{-k R})} + {\nu_{\pm}\over4}\log2 \eqno(29) $$
With this, we have:

$${\cal Z}_{\pm}= e^{-L
E_R^{(\pm)}}\bigl(g_{\pm}(R)e^{-R\varepsilon_{\lambda}}\bigr)^2\Sigma_{\pm}
\eqno(30)$$
Notice that $\Sigma_{\pm} \to 1$ as $L \to \infty$. $g_{\pm}$ can easily
be evaluated:

$$g_{+}(R)= {\sqrt{2\pi}\over\Gamma(\alpha + {1\over2})}\Big({\alpha\over
e}\Big)^\alpha \eqno(31)$$

$$g_{-}(R)= {2^{1\over4}\sqrt{\pi\alpha}\over{\Gamma(\alpha +
1)}}\big({\alpha\over e}\big)^\alpha \eqno(32)$$
where

$$\alpha = 2 h^2 R \eqno(33)$$
\vskip 0.2in

\noindent{\bf 4. Boundary States}
\vskip 0.1in

Now, we expect that:

$${\cal Z}_{\pm}= {\cal Z}_{h h} \pm {\cal Z}_{h -h}
\eqno(34)$$
where

$${\cal Z}_{h h}= \langle B_h | e^{-L H_R} | B_h\rangle \ ,\qquad {\cal
Z}_{h -h}=  \langle B_h | e^{-L H_R} | B_{-h}\rangle \eqno(35)$$
Here $|B_h\rangle$ is the boundary state corresponding to boundary
magnetic field $h$. In the large $L$ limit, the following expressions
are valid:

$${\cal Z}_{h h}= e^{-L E_R^{(+)}}\big(\langle 0| B_h\rangle\big)^2 +
e^{-L E_R^{(-)}}\big(\langle\sigma | B_h\rangle)^2 \eqno(36)$$

$${\cal Z}_{h -h}= e^{-L E_R^{(+)}}\big(\langle 0| B_h\rangle\big)^2 -
e^{-L E_R^{(-)}}\big(\langle\sigma | B_h\rangle)^2 \eqno(37)$$
In (36) and (37), $| 0\rangle$ is the ground state of energy $E_R^{(+)}$
of the Hilbert space ${\cal H}_R$ and it lies in the NS sector, while
$|\sigma\rangle$ is the lowest energy state of energy $E_R^{(-)}$ in the
R sector of ${\cal H}_R$: $|\sigma\rangle = \sigma\ |0\rangle$,\ $\sigma$
being the spin operator in the bulk theory. Also, in
the above, we have used

$$\langle 0| B_{-h}\rangle =  \langle 0| B_h\rangle \hbox{\qquad and
\qquad} \langle \sigma| B_{-h}\rangle = - \langle \sigma| B_h\rangle
\eqno(38)$$
Moreover we have chosen the phase of $|0\rangle$ to be such that
$\langle 0| B_h\rangle$ and $\langle \sigma | B_h\rangle$ are real (see
[17]). This can always be done. {}From (30), (34), (36), (37), it follows
that:

$$\langle 0| B_h\rangle=
{1\over\sqrt{2}}g_{+}(R)e^{-R\varepsilon_\lambda} \eqno(39)$$

$$\langle\sigma| B_h\rangle =
{1\over\sqrt{2}}g_{-}(R)e^{-R\varepsilon_\lambda} \eqno(40)$$

Now we construct the boundary states more explicitly. First we consider
the picture in which $x=time$ and $y=space$ (it is in this picture that
the boundary state $|B_h\rangle$ belongs to the Hilbert space
${\cal H}_R$) and write
down the mode expansions for the $\psi$ and ${\bar \psi}$ fermion
fields. In the Neveu-Schwarz and Ramond sectors we have
\vskip 0.2in
\noindent{\it Neveu-Schwarz sector :}

$$\psi(x,y)= \sum_{n=0}^{\infty}\big[a_{n +
{1\over2}}e^{-k_{n + {1\over2}}(x+i y)}+\
a^{\dagger}_{n + {1\over2}}e^{k_{n + {1\over2}}(x
+ i y)}\big] \eqno(41)$$

$${\bar \psi}(x,y)= \sum_{n=0}^{\infty}\big[{\bar a}_{n +
{1\over2}}e^{-k_{n + {1\over2}}(x- i y)} + {\bar a}^{\dagger}_{n +
{1\over2}}e^{k_{n + {1\over2}}(x-i y)}\big] \eqno(42)$$
where

$$k_{n + {1\over2}}= {2\pi (n +{1\over2})\over R}, \quad n \in {\bf Z}
\eqno(43)$$
Here $a^{\dagger}_{n + {1\over2}}$ ($a_{n + {1\over2}}$) and ${\bar
a}^{\dagger}_{n+ {1\over2}}$ (${\bar a}_{n+ {1\over2}}$) are the
creation (annihilation) operators for right moving and left moving
free Majorana fermions respectively. In the other sector we have:
\vskip 0.2in
\noindent{\it Ramond sector :}

$$\psi(x,y)= \sum_{n=0}^{\infty}\big[a_n e^{-k_n(x+ i y)} +
\ a^{\dagger}_n e^{k_n(x + i y)}\big] \eqno(44)$$

$${\bar \psi}(x,y)= \sum_{n=0}^{\infty}\big[{\bar a}_n e^{-k_n(x- i y)}
+\ {\bar a}^{\dagger}_n e^{k_n(x- i y)}\big] \eqno(45)$$
where

$$k_n = {2\pi n\over R} , \quad n \in {\bf Z} \eqno(46)$$
and with similar interpretation for the operators $a^{\dagger}_n$,
$a_n$, ${\bar a}^{\dagger}_n$, ${\bar a}_n$ as in the NS case.
\vskip 0.2in
\noindent In terms of these mode creation and annihilation operators, the
boundary
condition (4) can be expressed as the following set of constraints on any
boundary state $|B_h\rangle$:

$$a_l |B_h\rangle = i{k_l - \lambda\over k_l +
\lambda}{\bar a}^{\dagger}_l |B_h\rangle \eqno(47)$$
Here $l\in {\bf Z}+{1\over2}$ if the ${a_l}'s$ are in the NS sector and
$l\in {\bf Z}$ if the ${a_l}'s$ are in the R sector, and $k_l= {2 \pi
l\over R}$. {}From equations (39) and (40) and mode expansions (41) -- (46),
the boundary state takes the following form (following [11]):

$$\eqalign{|B_{\pm h}\rangle =
{1\over\sqrt{2}}g_{+}(R)e^{-R\varepsilon_{\lambda}}
\exp\Big\{\sum_{n=0}^{\infty}{{\cal K}(k_{n+1/2})\ a^{\dagger}_{n+1/2}\ {\bar
a}^{\dagger}_{n+1/2}}\Big\}|0\rangle\Big\}\ \pm\cr
\pm\ {1\over\sqrt{2}}g_{-}(R)e^{-R\varepsilon_{\lambda}}\
\exp\Big\{\sum_{n=1}^{\infty}{{\cal K}(k_n)\ a^{\dagger}_n\ {\bar
a}^{\dagger}_n}\Big\}|\sigma\rangle} \eqno(48)$$
where

$${\cal K}(k_l)= {\cal R}(i k_l)= i{k_l-\lambda\over k_l+\lambda} =
i{l-\alpha\over l + \alpha} \quad,\qquad \alpha=2 h^2 R \eqno(49)$$
where ${\cal R}(k)=i{k-i\lambda\over k+i\lambda}$ is the ``massless
boundary scattering matrix''
for our model [11]; ${\cal K}$ can also be obtained directly from (47).
Putting (31) -- (33), (48) and (49) together, we finally obtain the
boundary state:

$$\eqalign{|B_{\pm}\rangle = e^{-R\varepsilon_{\lambda}}\Big({\alpha\over
e}\Big)^{\alpha}\sqrt{\pi}\Big[{1\over{\Gamma(\alpha +
{1\over2})}}\exp
i\big\{\sum_{n=0}^{\infty}{{n+{1\over2}-\alpha}\over{n+{1\over2}+\alpha}}\
a^{\dagger}_{n+1/2}\ {\bar a}^{\dagger}_{n+1/2}\big\} |0\rangle\ \pm\cr
\pm\ {{2^{1\over4}\sqrt{\alpha}}\over{\Gamma(\alpha+1)}}\ \exp i\big\{\sum_{n=
1}^{\infty}\ {{n-\alpha}\over{n+\alpha}}\ a^{\dagger}_n\ {\bar
a}^{\dagger}_n\big\}
|\sigma\rangle\Big]} \eqno(50)$$
This expression (50) is the main result of this paper.

Now, we derive from (50) the boundary states corresponding to free and
fixed boundary conditions. These correspond to the UV and IR limits of
the boundary RG
flow as discussed earlier.
\vskip 0.2in
\noindent$\underline{\hbox{\it free case ($h = 0$) :}}$
\vskip 0.1in
\noindent {}From (50), we readily obtain, putting $\alpha=0$,

$$|B_{free}\rangle = \exp i\big\{\sum_{n=1}^{\infty}\ a^{\dagger}_{n
+{1\over2}}\ {\bar a}^{\dagger}_{n +{1\over2}}\big\}\ |0\rangle + i \exp
i\big\{\sum_{n=1}^{\infty}a^{\dagger}_{n +{1\over2}}\ {\bar
a}^{\dagger}_{n +{1\over2}}\big\}\ a^{\dagger}_{1\over2}\ {\bar
a}^{\dagger}_{1\over2}\ |0\rangle \eqno(51)$$
The above follows since the creation operators anticommute and since the
second term in (50) is zero in this case. Hence,

$$|B_{free}\rangle = \exp i\big\{\sum_{n=1}^{\infty}\
a^{\dagger}_{n+{1\over2}}\ {\bar a}^{\dagger}_{n+{1\over2}}\big\}\Big(|0\rangle
- |\varepsilon\rangle\Big) \eqno(52)$$
since

$$a^{\dagger}_{1\over2}\ {\bar a}^{\dagger}_{1\over2}\ |0\rangle = i\
|\varepsilon\rangle \eqno(53)$$
$\varepsilon$ being the energy density operator of the bulk critical
Ising model.
\vskip 0.2in
\noindent$\underline{\hbox{\it fixed case ($h \to \pm\infty$) :}}$
\vskip 0.1in
\noindent In this case, we have from (50):

$$\eqalign{|B_{fixed\pm}\rangle = {1\over\sqrt{2}}\exp
-i\big\{\sum_{n=1}^{\infty}\ a^{\dagger}_{n +{1\over2}}\ {\bar
a}^{\dagger}_{n +{1\over2}}\big\}\ \Big(|0\rangle +
|\varepsilon\rangle\Big)\ \pm\cr
\pm\ {1\over{2^{1\over4}}}\ \exp -i\big\{\sum_{n=1}^{\infty}\
a^{\dagger}_n\
{\bar a}^{\dagger}_n\big\}\ |\sigma\rangle } \eqno(54)$$

\vskip 0.1in
\noindent Both (52) and (54) agree with well known results of Cardy [9].
\vskip 0.4in
\noindent{\bf Conclusion}
\vskip 0.1in

The expression (50) for $|B_{\pm h}\rangle$ is the main result of this
paper. In the conformal limits, it matches with Cardy's results [9]. Our
method of
computing the boundary state is the same in spirit as Cardy's
[9,10] in deriving boundary states in boundary CFT -- namely, making use
of a (modular) transformation to express the partition function in one
channel in terms of states in the cross channel. An essentially similar
calculation can be done for the off critical Ising model with boundary
magnetic field; we will not deal with it here. It would be interesting
to extend the method to  general interacting boundary
integrable field theories.
\vskip 0.4in
\noindent{\bf Acknowledgement}
\vskip 0.1in
I am especially grateful to A.Zamolodchikov for his advice, inspiration
and numerous illuminating discussions all along. I am also thankful to
V.Brazhnikov for many helpful discussions.
\vfill\eject

\vskip 0.6in
\noindent{\bf References}
\vskip 0.1in

\item{1.} ``Conformal Invariance and Applications to Statistical
Mechanics'', eds. C.Itzykson, H.Saleur and J.B.Zuber (World Scientific,
1988).

\item{2.} A.Zamolodchikov, Advanced Studies in Pure Mathematics, 19 (1989)
641

\item{3.} C.N.Yang and C.P.Yang, J. Math. Phys. 10 (1969) 1115

\item{4.} E.H.Lieb and W.Liniger, Phys. Rev. 130 (1963) 1605

\item{5.} Al.Zamolodchikov, Nucl.Phys. B342 (1990) 695

\item{6.} T.Klassen and E.Melzer, Nucl. Phys. B350 (1991) 635

\item{7.} A.Zamolodchikov, Sov. J. Nucl. Phys. 46(6) (1987) 1090

\item{8.} J.Cardy, ``Conformal Invariance and Surface Critical Behavior'' in
Ref.1.

\item{9.} J.Cardy, Nucl. Phys. B324 (1989) 581

\item{10.} J.Cardy and D.Lewellen, Phys. Lett. B259 (1991) 274

\item{11.} S.Ghoshal and A.Zamolodchikov, Int. J. Mod. Phys. A9, No.21
(1994) 3841

\item{12.} S.Ghoshal,  Int. J. Mod. Phys. A9, No.27, (1994) 4801

\item{13.} S.Ghoshal, Phys. Lett. B334 (1994) 363

\item{14.} L.Chim, ``Boundary S Matrix for Integrable q-Potts model'',
RU-94-33, hep-\break th/9404118

\item{15.} A.Fring and R.Koberle, ``Boundary States in Affine Toda Field
Theory in the Presence of Reflecting Boundaries'', USP-IFQSC/TH/93-12,
hep-th/9304144

\item{16.} A.Fring and R.Koberle, ``Boundary Bound States in Affine Toda
Field Theory'', USP-IFQSC/TH/94-03, hep-th/9404188

\item{17.} I.Affleck and A.Ludwig'', Phys. Rev. Lett. 67 (1991) 161

\item{18.} R.Chatterjee and A.Zamolodchikov, Mod. Phys. Lett. A9, No. 24
(1994) 2227

\item{19.} P.Fendley and H.Saleur, ``Deriving Boundary S Matrices'',
USC-94-001, hep-\break th/9402045

\item{20.} P.Fendley, H.Saleur and N.P.Warner, ``Exact Solution of a
Massless Scalar Field With a Relevant Boundary Interaction'', USC-94-10,
hep-th/94-06125

\item{21.} B.M.McCoy and T.T.Wu, Phys. Rev. 162 (1967) 436

\item{22.} B.M.McCoy and T.T.Wu, Phys. Rev. 174 (1968) 546

\item{23.} M.M.McCoy and T.T.Wu, ``The Two--Dimensional Ising Model'',
Harvard University Press, (1973) Cambridge Mass.

\vfill\eject
\vskip 0.5in
\centerline{\hbox{\psfig{figure=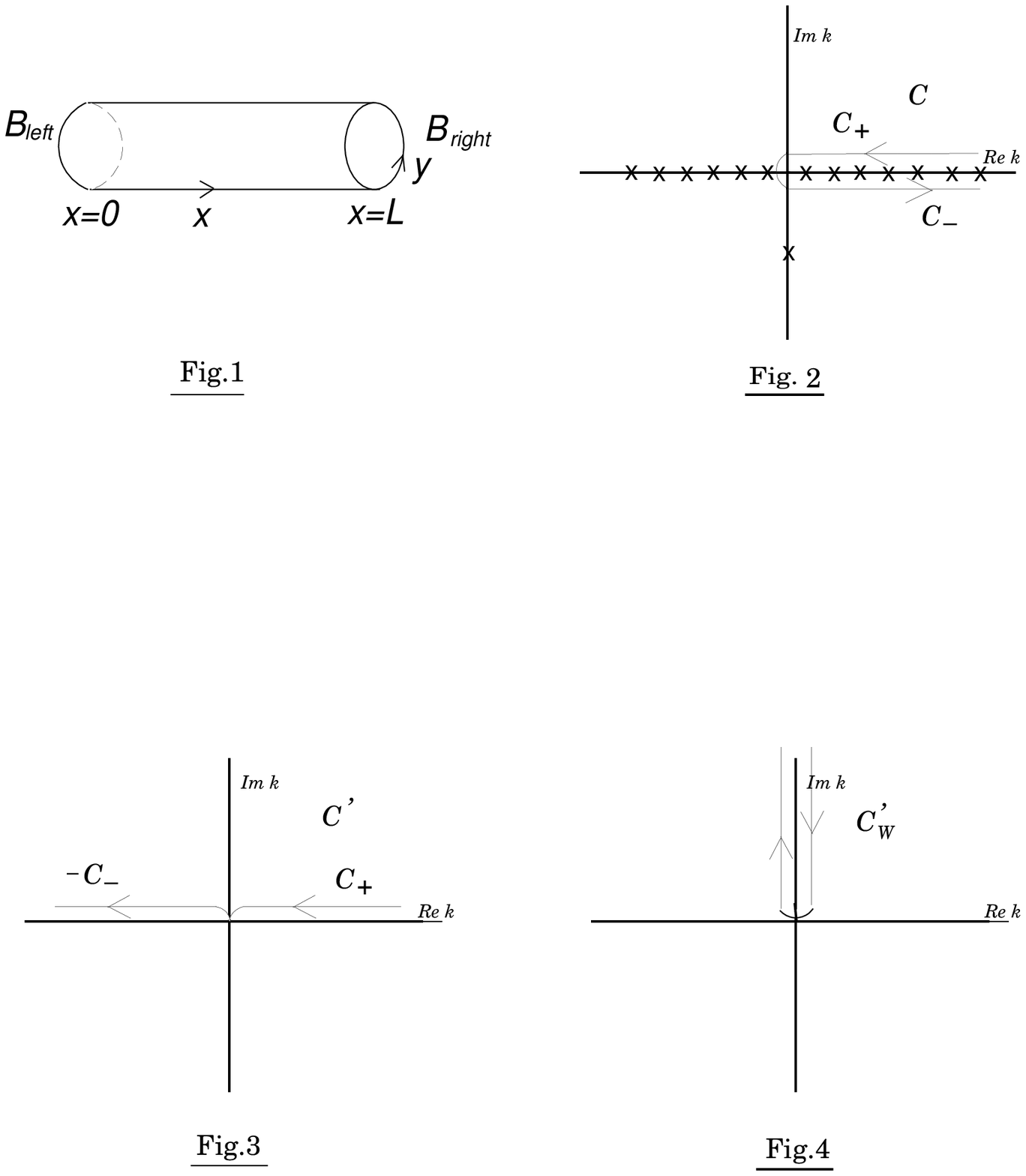}}}
\end